\documentclass[twocolumn,aps,prl,showpacs,tightenlines]{revtex4-1}
\usepackage{amsmath}
\usepackage{amsfonts}
\usepackage{graphicx}
\usepackage{epsfig}
\usepackage{color}
\usepackage{txfonts}
\usepackage[colorlinks,citecolor=blue]{hyperref}

\begin{document}

\title{Ultrastrong Jaynes-Cummings Model}
\author{Jin-Feng Huang}
\email{jfhuang@hunnu.edu.cn}
\affiliation{Key Laboratory of Low-Dimensional Quantum Structures and Quantum Control of
Ministry of Education, Department of Physics and Synergetic Innovation
Center for Quantum Effects and Applications, Hunan Normal University,
Changsha 410081, China}
\author{Jie-Qiao Liao}
\email{jqliao@hunnu.edu.cn}
\affiliation{Key Laboratory of Low-Dimensional Quantum Structures and Quantum Control of
Ministry of Education, Department of Physics and Synergetic Innovation
Center for Quantum Effects and Applications, Hunan Normal University,
Changsha 410081, China}
\author{Le-Man Kuang}
\affiliation{Key Laboratory of Low-Dimensional Quantum Structures and Quantum Control of
Ministry of Education, Department of Physics and Synergetic Innovation
Center for Quantum Effects and Applications, Hunan Normal University,
Changsha 410081, China}
\date{\today}

\begin{abstract}
We propose a reliable scheme to realize the ultrastrong Jaynes-Cummings (JC) model by simultaneously modulating the resonance frequencies of the two-level system and the bosonic mode in the ultrastrong quantum Rabi model. We find that in both the high- and low-frequency modulation regimes, the counter-rotating terms can be completely suppressed without reducing the coupling strength of the rotating-wave terms, and hence the ultrastrong JC Hamiltonian is achieved. The ultrastrong JC interaction can not only be used to implement ultrafast quantum operations, but also will open up a new route to the demonstration of quantum phase transition associated with the JC Hamiltonian across the deep-strong coupling point. Some discussions on the experimental implementation of this scheme with circuit-QED systems are presented.
\end{abstract}
\maketitle

\emph{Introduction}---The ultrastrong-coupling (USC) regime is a new realm for understanding the light-matter interactions in the frontier of physics, ranging from quantum optics to condensed matter physics and quantum information science~\cite{Solanoreview,Norireview}. In the USC regime, the light-matter coupling strength reaches an appreciable fraction of the resonance frequencies of the subsystems, which leads to the significance of the counter-rotating (CR) interaction terms. This feature not only motivates the study of the integrability of related physical models~\cite{Braak2011PRL,Chen2012PRA,Lee2013JPA,Xie2014PRX}, but also stirs up the studies of various CR-interaction-caused physical effects~\cite{Klimov2009book,Zheng2008PRL,Blais2009PRA,Ciuti2010PRL,Solano2010PRL,Ashhab2010PRA,Ai2010PRA,Ciuti2011PRL,Hartmann2012PRL,Hartmann2013PRL,Savasta2013PRL,Law2013PRA,Huang2014,Moreno2014PRL,Garziano2015,Liu2017PRL,Shi2018PRL} and the manipulation of the CR terms~\cite{huang1,huang2}. In experiments, the USC regime has been demonstrated in various physical platforms, including semiconductor cavity-QED systems~\cite{Beltram2009PRB,Huber2009Nat,Sirtori2010PRL,Faist2012}, superconducting circuit-QED systems~\cite{Gross2010NatPhy,Mooij2010PRL,Semba2017,Lupascu2017,You2017,Steele2017}, coupled photon-2D-electron-gas~\cite{Faist2012Sci,Wegscheider2014,Kono2016}, light-molecule~\cite{Ebbesen2011,Gigli2014}, and photon-magnon systems~\cite{Tobar2014}.

In this Letter, we propose a new parameter regime of the light-matter interaction: the ultrastrong Jaynes-Cummings (JC) model, which describes the coupling between a single-mode bosonic field (oscillator, resonator, cavity field etc.) and a two-level system (TLS, natural or artificial two-state atom or qubit). This ultrastrong JC model possesses the JC-type interaction and works in the ultrastrong-coupling regime, which correspond to the two advantages of excitation conservation and ultrafast quantum operation, respectively. The feature of excitation conservation is an important element for realization of some high-performance quantum tasks based on excitation exchange such as quantum state transfer. Concretely, we propose a reliable scheme to completely suppress the CR terms in the quantum Rabi model without modifying the rotating-wave terms in the USC regime. Hence, an ultrastrong JC-type interaction between a TLS and a single bosonic mode is obtained. This ultrastrong JC model will shed light on the study of ultrafast quantum operation~\cite{Solano2012PRL} and quantum phase transition in the JC Hamiltonian when the coupling strength equals one of the resonance frequencies of the TLS and the bosonic mode.

\emph{Model}---Let us consider the quantum Rabi model which describes the interaction between a TLS and a single bosonic mode. The Hamiltonian of the quantum Rabi model reads~\cite{Rabi1936,Rabi1937}
\begin{equation}
H_{\text{Rabi}}=H_{\text{JC}}+H_{\text{CR}},
\end{equation}
with
\begin{subequations}
\label{eq:HR}
\begin{align}
H_{\text{JC}} & =\frac{\omega_{0}}{2}\sigma_{z}+\omega_{c}a^{\dagger}a+g(a\sigma_{+}+a^{\dagger}\sigma_{-}),\label{ultJCH}\\
H_{\text{CR}} & =g(a^{\dagger}\sigma_{+}+a\sigma_{-}),
\end{align}
\end{subequations}
where $H_{\text{JC}}$ is the JC Hamiltonian~\cite{JC1963,Knight1993JMO} and $H_{\text{CR}}$ describes the CR interaction
terms. In Eq.~(\ref{eq:HR}), $\omega_{0}$ is the transition frequency of the TLS described by the
Pauli operator $\sigma_{z}=|e\rangle\langle e|-|g\rangle\langle g|$ and the raising and
lowering operators $\sigma_{+}=\sigma^{\dagger}_{-}=\vert e\rangle \langle g\vert$, $\omega_{c}$ is the resonance frequency of the bosonic mode described by the annihilation
and creation operators $a$ and $a^{\dagger}$, and $g$ denotes the Rabi coupling strength.

To manipulate the interactions in the quantum Rabi model, we apply
a pair of sinusoidal frequency modulations to the TLS and the bosonic
mode. The modulation Hamiltonian is given by
\begin{equation}
H_{M}(t)=\xi\nu\cos(\nu t)(\sigma_{z}/2+a^{\dagger}a),
\end{equation}
where $\xi$ and $\nu$ are the scaled modulation amplitude and modulation frequency,
respectively. For convenience, we express the total Hamiltonian as $H(t)=\omega_{a}(t)\sigma_{z}/
2+\omega_{c}(t)a^{\dagger}a+g(\sigma_{+}+\sigma_{-})(a+a^{\dagger})$,
with $\omega_{a}(t)=\omega_{0}+\xi\nu\cos(\nu t)$
and $\omega_{c}(t)=\omega_{c}+\xi\nu\cos(\nu t)$. In a rotating frame
defined by the unitary transformation operator $\exp\left\{-i[\omega_{c}t+\xi\sin(\nu t)]a^{\dagger}a-i[\omega_{0}t+\xi\sin(\nu t)]\sigma_{z}/2\right\}$,
the Hamiltonian $H(t)$ becomes $H_{1}(t)=H_{\text{JC}}^{I}+\hat{\varepsilon}(t)$,
where $H_{\text{JC}}^{I}=g(\sigma_{+}ae^{i\delta t}+a^{\dagger}\sigma_{-}e^{-i\delta t})$ with
$\delta=\omega_{0}-\omega_{c}$ is the JC Hamiltonian in the interaction picture
with respect to $\omega_{0}\sigma_{z}/2+\omega_{c}a^{\dagger}a$. Under this transformation, the CR
terms become
\begin{equation}
\hat{\varepsilon}(t)=g(\sigma_{+}a^{\dagger}e^{i(\omega_{0}+\omega_{c})t}e^{i2\xi\sin(\nu t)}+\mbox{H.c.}),\label{eq:E}
\end{equation}
which are expected to be neglected under proper parameter conditions.

Below, we propose two parameter regimes in which the CR terms $\hat{\varepsilon}(t)$ can be
ignored. (i) The high-frequency modulation regime: $\nu>\omega_{0}+\omega_{c}$. Using the Jacobi-Anger identity $\exp[i2\xi\sin(\nu t)]=\sum_{n=-\infty}^{\infty}J_{n}(2\xi)e^{in\nu t}$, the CR terms can then be expressed as $\hat{\varepsilon}(t)=g(\sigma_{+}a^{\dagger}\sum_{n=-\infty}^{\infty}J_{n}(2\xi)e^{i\Delta_{n}t}+\mbox{H.c.})
$ with the oscillating frequencies $\Delta_{n}=\omega_{0}+\omega_{c}+n\nu$ and the
Bessel function of the first kind $J_{n}(2\xi)$. We denote the index of the sideband with the smallest oscillating frequency as $n=n_{0}$, namely $|\Delta_{n_{0}}|=\text{min}\{|\Delta_{n}|=|\omega_{0}+\omega_{c}+n\nu|,n\in Z\}$, with $``Z"$ being the set of all integers. Under the parameter conditions $\nu\gg|\Delta_{n_{0}}|$, $\nu>\omega_{0}+\omega_{c}$,
and $\nu\gg g|J_{n}(2\xi)\sqrt{N}|$ with $N$ being the largest excitation number involved in
mode $a$, the CR terms for $n\neq n_{0}$ can be discarded with rotating-wave approximation (RWA) and then we obtain
$\hat{\varepsilon}(t)\simeq g_{c}(\sigma_{+}a^{\dagger}e^{i\Delta_{n_{0}}t}+\mbox{H.c.})$
with $g_{c}=gJ_{n_{0}}(\xi)$. We further choose a suitable $\xi$ and $\nu$ to grantee $g/
\delta\gg g_{c}/\Delta_{n_{0}}$ and $|g_{c}\sqrt{N}|\ll\Delta_{n_{0}}$,
then the CR Hamiltonian $\hat{\varepsilon}(t)$ can be safely ignored.
(ii) The low-frequency modulation regime: $\nu/\omega_{0}\ll1$. In
this regime, we consider the time period $\nu t\ll1$, the CR terms
become $\hat{\varepsilon}(t)\approx g(\sigma_{+}a^{\dagger}e^{i(\omega_{0}+\omega_{c}
+2\xi\nu)t}+\mbox{H.c.})$ under the expansion $\sin(\nu t)\approx\nu t$.
If we further choose a suitable $\xi$ to satisfy the parameter conditions
$\omega_{0}+\omega_{c}+2\xi\nu\gg g\sqrt{N}$ and $\delta<\omega_{0}$,
then the CR terms $\hat{\varepsilon}(t)$ can be safely discarded.

\begin{figure}
\includegraphics[bb= 12 250 526 717, width=0.475\textwidth]{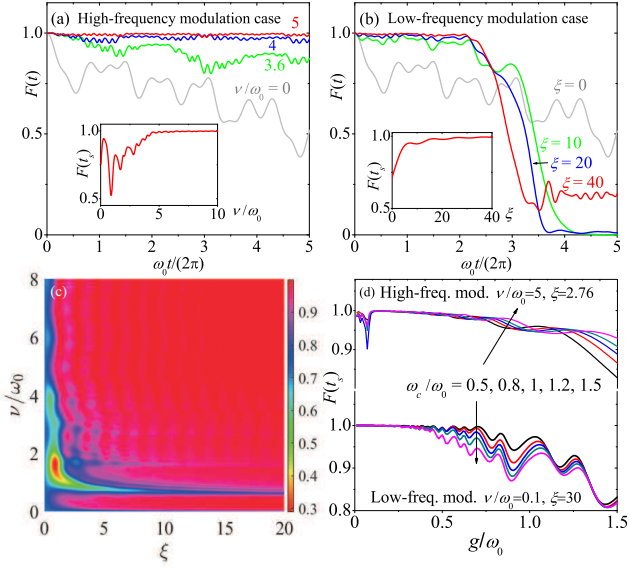}
\caption{(Color online) Dynamics of the fidelity $F(t)$ in the (a) high-frequency
and (b) low-frequency modulation cases. In panel (a), the parameters
are taken as $\nu/\omega_{0}=3.6$, $4$, and $5$, and $\xi=2.76$ corresponding
to $J_{0}(2\xi)=0$. In panel (b), we take $\nu/\omega_{0}=0.1$ and
$\xi=10$, $20$, $40$. We also present the fidelity (gray curves) corresponding
to the unmodulated case ($\nu/\omega_{0}=0$ or $\xi=0$) for comparison.
In these two insets, the fidelity $F(t_{s})$ at time $t_{\textrm{s}}=\pi/(2g)$
is plotted versus either $\nu$ or $\xi$. (c) The fidelity $F(t_{s})$
versus the two tunable modulation parameters $\nu$ and $\xi$. (d)
$F(t_{s})$ versus the ratio $g/\omega_{0}$ in both the high-frequency
($\nu/\omega_{0}=5$, $\xi=2.76$) and the low-frequency ($\nu/\omega_{0}=0.1$,
$\xi=30$) modulation cases when the frequencies of the TLS and the
bosonic mode take different values. The initial state of the system
is $(|g\rangle+|e\rangle)|\alpha\rangle/\sqrt{2}$ with $\alpha=0.1$,
and other parameters used in panels (a-c) are given by $g/\omega_{0}=0.5$
and $\omega_{c}=\omega_{0}$.}
\label{Fig1}
\end{figure}

\emph{Fidelity}---To evaluate the validity of the RWA made in
the derivation of the Hamiltonian $H_{\text{JC}}^{I}$, we check the fidelity $F(t)=|
\langle\phi(t)|\psi(t)\rangle|^{2}$ between the exact state $|\phi(t)\rangle$ governed by the
exact Hamiltonian $H_{1}(t)$ and the approximate state $|\psi(t)\rangle$ governed by
the ultrastrong JC Hamiltonian $H_{\text{JC}}^{I}$. Without loss of generality,
we assume the initial state $|\phi(0)\rangle=|\psi(0)\rangle=(1/\surd2)(|g\rangle+|e\rangle)|
\alpha\rangle$ of the system and calculate the fidelity in both the high- and low-frequency
modulation cases. For the high-frequency modulation case, in Fig.~\ref{Fig1}(a) we plot
the fidelity $F$ as a function of time $t$ when the modulation frequency
takes various values: $\nu/\omega_{0}=0$, $3.6$, $4$, and $5$. Here
we consider the resonant case $\omega_{0}=\omega_{c}$ and choose
the scaled modulation amplitude as $\xi=2.76$ such that $J_{0}(2\xi)=0$. These plots show that
a higher fidelity can be obtained for a larger value of the modulation
frequency $\nu/\omega_{0}$, keeping in consistent with the parameter
condition for the approximation. This feature can also be seen from the inset plot, which shows the envelop of the fidelity
at time $t_{s}=\pi/(2g)$ as an increasing function of the parameter $\nu/\omega_{0}$.

In the low-frequency modulation case, we plot in Fig.~\ref{Fig1}(b) the dynamics of the
fidelity when the modulation amplitude $\xi$ takes various values: $\xi=0$,
$10$, $20$, and $40$. Here we can see that the fidelity is higher for
a larger modulation amplitude $\xi$ within the first two Rabi oscillation cycles [$\omega_{0}t/
(2\pi)<2$] for $\nu=0.1\omega_{0}$. Since the Rabi oscillation frequency $g$ ($g\sim\omega_{0}$ is possible
in the ultrastrong-coupling regime) is much larger than $\nu$ when
$\nu/\omega_{0}\ll1$, the system can experience several Rabi oscillations in a short time duration
$ t\ll1/\nu$. We also checked that the fidelity $F(t_{s})$ at time $t_{s}$ increases with the increase of the value $\xi\nu$,
as explained by the parameter conditions in the lower-frequency modulation case.

We further plot the fidelity $F(t_{s})$ as a function of the two tunable modulation
parameters $\nu/\omega_{0}$ and $\xi$. As shown in Fig.~\ref{Fig1}(c), the fidelity is almost $1$
in two regions, which correspond to the high- and low-frequency modulation regimes.
In the high-frequency modulation regime, the fidelity is high when the corresponding
parameter conditions are satisfied. Here we can see that the fidelity
oscillates slightly with the parameter $\xi$. This oscillation is
caused by the frequency modulation, which can be seen from the facts
that the oscillation period of the fidelity matches that of the Bessel
function $\vert J_{0}(2\xi)\vert$, which determines the coupling
strength of the CR terms, and that the peaks (dips) of the fidelity
correspond to the dips (peaks) of the function $\vert J_{0}(2\xi)\vert$.

Though we take the zero points of the Bessel function $J_{0}(2\xi)$ to eliminate the CR interaction terms
in the idea case, our scheme works in a wide range of $\xi$ because
the relation $\vert J_{0}(2\xi)\vert\ll1$ is established in a wide range, and hence the CR terms can be
neglected when $\omega_{0}+\omega_{c}\gg gJ_{0}(2\xi)$. In the low-frequency modulation
regime, the fidelity is almost $1$ when the corresponding parameter conditions are satisfied. Here
we can see that the fidelity $F(t_{s})$ increases with the increasing of the parameter $\xi$. In panel (d), we also
plot the fidelity $F(t_{s})$ as a function of $g/\omega_{0}$ when $\omega_{c}/\omega_{0}$ takes various values. Here, 
the fidelity decreases slightly when the coupling strength is larger than half of the resonance frequency $\omega_{c}$.
However, the fidelity is still larger than $0.8$ even when the system
enters the deep-strong-coupling regime (until $g/\omega_{0}\sim1.5$).

\emph{Ultrafast Rabi oscillation and state transfer}---
Assume that the system is initially in state $|g,0\rangle$, which
is the ground state of the JC Hamiltonian when $g/\omega_{0}<1$ in
the resonant case $\omega_{0}=\omega_{c}$. When the CR
terms are completely suppressed, the system is described by
the JC Hamiltonian and it will always stay in state $|g,0\rangle$. The deviation of the state $|g,0\rangle$
can be used to evaluate the validity of the approximate Hamiltonian $H_{\text{JC}}^{I}$
because the deviation of the population is caused by the CR
terms. On the other hand, when the system is initially in state
$|e,0\rangle$, then the JC Hamiltonian will govern the Rabi oscillation
between the two states $|e,0\rangle$ and $|g,1\rangle$, and the
deviation of the Rabi oscillation is caused by the CR terms.

\begin{figure}
\includegraphics[bb= 1 20 532 478, width=0.475\textwidth]{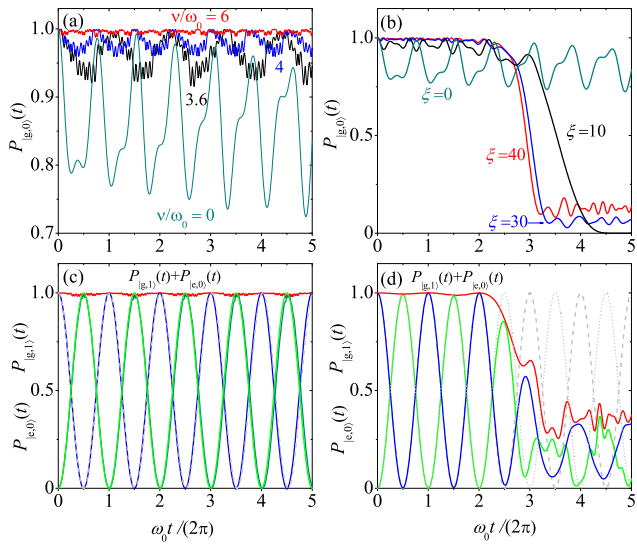}
\caption{(Color online) Dynamics of the population $P_{\vert g,0\rangle}(t)$ in the (a) high-frequency
and (b) low-frequency modulation cases when
the system starts from the state $\vert g,0\rangle$. In panel (a),
the modulation frequency is taken as $\nu/\omega_{0}=0$, $3.6$,
$4$, and $6$, the modulation amplitude is taken as $\xi=2.76$,
corresponding to a zero value of the Bessel function $J_{0}(2\xi)$.
In panel (b), $\xi=0$, $10$, $30$, and $40$ are taken in the low-frequency
modulation case $\nu/\omega_{0}=0.1$. In panels (c) and (d), the
populations $P_{\vert e,0\rangle}(t)$ (blue curve) and $P_{\vert g,1\rangle}(t)$
(green curve) are plotted in the high- and low-frequency modulation
cases when the system starts from the state $|e,0\rangle$. The exact
numerical results are compared to the Rabi oscillation (gray curves)
determined by the JC Hamiltonian. In panel (c), the
parameters are taken as $\xi=2.76$ and $\nu/\omega_{0}=5$, while
in panel (d) these are taken as $\xi=40$ and $\nu/\omega_{0}=0.1$.
Other parameters are given by $\omega_{c}=\omega_{0}$ and $g/\omega_{0}=0.5$.}
\label{Fig2}
\end{figure}

In Figs.~\ref{Fig2}(a) and~\ref{Fig2}(b), we show the population
dynamics of the state $|g,0\rangle$ corresponding to the high- and
low-frequency modulation cases when the system is
in the initial state $|g,0\rangle$. In the ultrastrong-coupling regime, the probability $P_{|g,0\rangle}$ of the system in the absence of modulation is oscillatory and deviates from $1$ significantly. Under the
high-frequency modulation condition [Fig~\ref{Fig2}(a)], a smaller
deviation is obtained for a larger modulation frequency, which is
in consistent with the parameter conditions of the RWA. In the low-frequency
modulation case [Fig~\ref{Fig2}(b)], the deviation is smaller
for a larger modulation amplitude $\xi$. This is because the validity
of the JC Hamiltonian is better for a larger value of $\xi\nu$. In
particular, it should be emphasized that the validity of the JC Hamiltonian
in the low-frequency modulation case is only established in the short-time
limit $t\ll1/\nu$, as confirmed by Fig.~\ref{Fig2}(b). However, the validity period of the ultrastrong JC
Hamiltonian could still be longer than the time scales $1/\omega_{c}$ and $1/\omega_{0}$
because of $\nu\ll{\omega_{c},\omega_{0}}$.

Corresponding to the initial state $|e,0\rangle$, the system governed by the JC
Hamiltonian will experience a Rabi oscillation between the two states
$|e,0\rangle$ and $|g,1\rangle$. In Figs.~\ref{Fig2}(c) and~\ref{Fig2}(d),
we show the exact evolution of the system populations $P_{|e,0\rangle}(t)$
(blue curve) and $P_{|g,1\rangle}(t)$ (green curve) under the high- and
low-frequency modulations, respectively. We also present the
Rabi oscillation for comparison (gray curves). Here we can see that
the system transits between the two states $|e,0\rangle$ and $|g,1\rangle$
following the Rabi oscillation. This means that the system under the modulation can
be well described by the JC Hamiltonian. We also find that the JC Hamiltonian describes the
system well only during about two Rabi oscillation periods in the low-frequency modulation case.
In addition, the period of the Rabi oscillation is $\pi/g$, which
could be much shorter than the Rabi oscillation period in the conventional
JC Hamiltonian case because the coupling strength $g$ in the present
case corresponds to the ultrastrong coupling regime. When we choose
a deep-strong coupling case $g/\omega_{0}>1$, an ultrafast Rabi oscillation
can be implemented in the sense that the oscillation is faster than
the free evolution of the TLS and the bosonic mode. In the
JC regime, the total population in the single-excitation subspace is a conserved quantity,
which can be seen from the normalization of the populations in the single-excitation subspace.

An interesting application of the ultrafast Rabi oscillation in this
model is ultrafast quantum state transfer between the bosonic mode and
the TLS. Consider an initial state $|\psi(0)\rangle=(1/\surd2)(|g\rangle+|e\rangle)|0\rangle$,
after an evolution duration $t_{\textrm{s}}\sim1/g$, the state becomes $|\psi(t_{\textrm{s}})
\rangle=|g\rangle\otimes(1/\surd2)(|0\rangle-i|1\rangle)$, i.e., a transfer of the superposition
information from the TLS to the bosonic mode. This means an ultrafast quantum state transfer because the operation time could
be shorter than one free-evolution period of the TLS and the bosonic mode. To estimate the
performance of the quantum state transfer, we numerically calculate the fidelity between the
target state corresponding to the JC Hamiltonian and the transferred state governed by the exact Hamiltonian. By investigating
the fidelity as a function of the two modulation parameters $\nu/\omega_{0}$
and $\xi$, we find that the fidelity is almost $1$ under the parameter conditions of the RWA and has
a similar behavior as the fidelity in Fig.~\ref{Fig1}(c).

\begin{figure}
\includegraphics[bb= 6 3 525 245, width=0.475\textwidth]{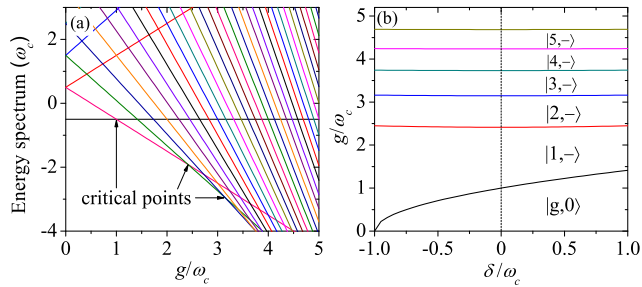}
\caption{(Color online) (a) Energy spectrum of the ultrastrong JC model at resonance $\omega_{0}=\omega_{c}$ as a function of the coupling strength $g/\omega_{c}$. (b) Phase diagram of the ground state of the ultrastrong JC model as a function of the detuning $\delta/\omega_{c}$ and the coupling strength $g/\omega_{c}$.}
\label{Fig3}
\end{figure}

\emph{Quantum phase transition}---The ultrastrong JC interaction opens up a new route to study the
quantum phase transition. When the coupling strength $g$ sweeps through the critical point
$g/\omega_{0}=1$ in the resonant case $\omega_{0}=\omega_{c}$, the ground state of the
JC model changes from $|g,0\rangle$ to the eigenstate $|1-\rangle=(\vert g,1\rangle-\vert e,
0\rangle)/\sqrt{2}$ in the single-excitation subspace, as shown in Fig.~\ref{Fig3}(a). This indicates that a quantum phase
transition can occur in the ultrastrong JC Hamiltonian. This effect can only be observed in the proposed ultrastrong JC
model, because the JC-type Hamiltonian and the critical-point-cross
parameter condition can not be satisfied simultaneously in the conventional JC model. We note that the signature of this
quantum phase transition can be seen from the excited state population $P_{\vert e\rangle}$
of the TLS. When one turns the coupling strength $g$ through the critical point, the
population $P_{\vert e\rangle}$ will step change from $0$ to $1/2$. This signature can be
detected by measuring the population of the TLS.

When the coupling strength $g$ keeps increasing, the ground state of the ultrastrong JC Hamiltonian changes following the order $\vert 1,-\rangle\rightarrow\vert 2,-\rangle\rightarrow\vert 3,-\rangle\cdots$. This is because the eigenenergy of the eigenstate $\vert n,-\rangle$ in the resonant case $\omega_{0}=\omega_{c}$ contains the term $-g\sqrt{n}$. With the increase of $g$, the ground state will be changed from state $\vert n,-\rangle$ to $\vert n+1,-\rangle$ gradually [Fig.~\ref{Fig3}(a)]. In Fig.~\ref{Fig3}(b), we show the phase boundary of the ground state as a function of the detuning $\delta/\omega_{c}$ and the coupling strength $g/\omega_{c}$. Here, the location of the cross points at the resonant case $\delta=0$ in panel (b) corresponds to the critical points in panel (a).

\emph{Discussions on the experimental implementation}---The present modulation method is general and it can be implemented with various ultrastrongly-coupled quantum systems,
which could be described by the quantum Rabi model, such as superconducting circuit-QED systems~\cite{Semba2017,Gu2018PRep}, various semiconductor~\cite{Faist2012}, organic molecule~\cite{Gigli2014} and cavity-QED systems~\cite{Tobar2014}.
Below we focus our discussions on the circuit-QED setup, in which the deep-strong coupling regime has been observed
with $g/\omega_{c}=1.34$~\cite{Semba2017}. Concretely, the resonance frequencies of the TLS and the bosonic mode could
be of the order of $\omega_{a}=\omega_{c}\sim2\pi\times3$ - $10$ GHz, the coupling strength $g$ could enter the USC regime and even the deep-strong coupling regime. The Rabi oscillation period is $T\sim10^{-10}$-$10^{-9}$ s, which is much shorter than the Rabi oscillation period in a typical circuit-QED system with $g$ at the order of several megahertz and $T\sim10^{-6}$ s. The two modulation parameters $\xi$ and $\nu$ can be tuned by proper controlling the biasing signal of the qubit and the resonator. The frequency modulation of the transmission line resonator can be realized by introducing a SQUID boundary to the resonator and changing the flux through the loop of the SQUID~\cite{Wallquist2006PRB,Sandberg2008APL,Yamamoto2008APL,Johansson2009PRL,Wilson2010PRL,Wilson2011Nature}. The frequency modulation of the qubit can be realized by introducing a longitudinal driving to the qubit~\cite{Mooij2009,Porras2014PRL,Porras2012PRL}. In the low-frequency modulation case, a fidelity $F>0.94$ can be obtained when $\xi\nu/\omega_{0}\sim0.5$ with $g/\omega_{0}=0.5$. A smaller driving amplitude $\xi\nu/\omega_{0}\sim0.1$ also works ($F\sim0.99$) when $g/\omega_{0}=0.1$ is used in simulation.
For the high-frequency modulation case, an optimal driving amplitude can be chosen by taking a small work value of $\xi$. Therefore, this modulation scheme should be within the reach of current and the near-future experimental techniques.

\emph{Conclusion}---In conclusion, we proposed a new physical regime of the light-matter interaction: an ultrastrong JC model. We also studied the implementation of ultrafast quantum state transfer and quantum phase transition in this ultrastrong JC model. This study will not only enrich the form and parameter regime of the light-matter interactions, but also widen the potential applications of circuit-QED systems in modern quantum technology.

\emph{Acknowledgments.}---The authors thank Prof. Cristiano Ciuti and Prof. Chi Kwong Law for helpful discussions. J.-F.H. is supported by the NSFC Grant No.~11505055. J.-Q.L.
is supported in part by NSFC Grants No.~11822501 and No.~11774087, and HNNSFC Grant
No.~2017JJ1021. L.M.K. is supported by the National Natural Science
Foundation of China under Grants No. 11775075 and No. 1143401.

\end{document}